# Unravelling the Role of Stacking Disorder on the Optoelectronic Properties of $Zn_3P_2$

Francesco Salutari[1,†], Nico Kawashima[2,3,†], Aidas Urbonavicius[4], Helena Rabelo Freitas[1], Raphael Lemerle[5], Thomas Hagger[5], Kimberly A. Dick[4], Anna Fontcuberta i Morral[5,6], Simon Escobar Steinvall[4], Maria Chiara Spadaro[1,7,8], Silvana Botti[2,*], Jordi Arbiol[1,9,*]

[1]Catalan Institute of Nanoscience and Nanotechnology – ICN2 (CSIC and BIST), Campus UAB, Bellaterra, Barcelona, 08193, Catalonia, Spain
[2]Research Center Future Energy Materials and Systems and Interdisciplinary Centre for Advanced Materials Simulation, Faculty of Physics and Astronomy, Ruhr University Bochum, Universitätsstraße 150, 44801 Bochum, Germany
[3]Institute of Condensed Matter Theory and Optics, Friedrich-Schiller-Universität Jena, Max-Wien-Platz 1, 07743 Jena, Germany
[4]Center for Analysis and Synthesis and NanoLund, Lund University, Box 124, 221 00 Lund, Sweden
[5]Laboratory of Semiconductor Materials, Institute of Materials, School of Engineering, Ecole Polytechnique Fédérale de Lausanne, 1015, Lausanne, Switzerland
[6]Institute of Physics, EPFL, Route Cantonale 1, Lausanne, 1015, Vaud, Switzerland
[7]Physics and Astronomy Department "Ettore Majorana" (DFA), Catania University, Via S. Sofia 64, Catania, 95123, Sicily, Italy
[8]IMM-CNR, Catania University Building, Via S. Sofia 64, Catania, 95123, Sicily, Italy
[9]ICREA, Pg. Lluıs Companys 23, Barcelona, 08010, Catalonia, Spain

†Equal Contribution
*Corresponding authors

## Abstract

Zinc phosphide ($Zn_3P_2$) is a promising photovoltaic absorber for thin-film and flexible solar cells due to its earth-abundant composition and favourable optoelectronic properties. Recent advances in epitaxy have enabled the growth of high-quality $Zn_3P_2$ thin films despite the challenges posed by its incompatible lattice parameter and thermal expansion coefficient. However, $Zn_3P_2$ remains prone to intrinsic extended defects, such as rotated domains, that can limit device performance. Here, using (scanning) transmission electron microscopy, we identify a previously unreported class of extended defects that appear as planar faults described by displacement vectors lying in the (001) plane. Within a pseudo-cubic description of $Zn_3P_2$, we establish a direct correspondence between planar faults and rotated domains, showing that both arise from the flexible ordering of vacant sites in the Zn sublattice. First-principles calculations reveal an extremely low planar-defect formation energy of 2.5 mJ $m^{-2}$, demonstrating that these defects form at essentially negligible energetic cost, in excellent agreement with their high experimentally observed occurrence. Additional density functional theory (DFT) calculations show that intrinsic planar defects neither introduce mid-gap electronic states nor significantly perturb the local electrostatic potential, indicating that they are electronically benign. Instead, we propose that planar defects indirectly degrade device performance by acting as preferential segregation sites for optically active point defects.

# 1. Introduction

The growing global demand for sustainable energy has intensified research into next-generation photovoltaic (PV) technologies that can overcome the intrinsic limitations of established systems. Crystalline silicon (c-Si), despite its technological maturity and widespread deployment, is fundamentally constrained by its indirect bandgap,[1] which leads to relatively weak optical absorption compared with direct bandgap semiconductors.[2] This limitation restricts its efficiency in thin-film or flexible device architectures, unless supplemented by sophisticated light-management strategies.[3] Furthermore, c-Si manufacturing relies on energy-intensive, high-temperature processes, contributing to extended energy payback times and a substantial embodied carbon footprint.[4] Alternative commercial thin-film technologies, including CdTe and $Cu(In,Ga)Se_2$ (CIGS), partially mitigate the optical limitations of silicon. However, CdTe-based devices depend on cadmium and tellurium, elements associated with toxicity concerns and limited availability,[5] while CIGS technologies rely on indium and gallium, both of which are subject to significant supply constraints and geopolitical risk.[6]

In this context, zinc phosphide ($Zn_3P_2$) has emerged as a promising photovoltaic absorber for thin-film and flexible solar cells.[7–11] This material combines the advantages of earth-abundant and non-toxic constituent elements with favorable optoelectronic properties,[12] including a direct bandgap of ~1.5 eV,[13] near optimal for single-junction devices,[1] strong optical absorption in the visible spectrum ($10^4$–$10^5$ $cm^{-1}$),[14] and long minority carrier diffusion lengths.[15,16]

During growth, $Zn_3P_2$ crystallizes in its α-phase, adopting a tetragonal structure (space group $P4_2/nmc$; a = 8.089 Å, c = 11.39 Å).[17,18] The primitive unit cell contains 40 atoms, comprising 16 P anions and 24 Zn cations, arranged in a layered configuration with Zn planes stacked along the c-axis and interleaved by P layers.[18] Each Zn atom is tetrahedrally coordinated by four P atoms, whereas every P atom sits at the center of a cube with 6 of the 8 corners occupied by Zn atoms.[19] This arrangement results in a cation sublattice with one-quarter of the sites vacant, consistent with the stoichiometry of the compound. Under the idealized condition $c = \sqrt{2}a$ and assuming atoms occupy their ideal lattice positions, the structure can be mapped onto a pseudo-cubic framework. Atomic ordering at interfaces with III–V semiconductors is more readily described within the pseudo-cubic framework, where the vacant sites ordering is neglected, and a reduced lattice parameter of 5.72 Å can be observed.[8,20–22] Moreover, the pseudo-cubic representation reduces computational complexity, enabling tractable band structure calculations despite the large primitive unit cell.[23–25]

Extended intrinsic defects in $Zn_3P_2$, such as rotated domains, are most naturally described within the pseudo-cubic framework. As first reported by Steinvall *et al.*[26] and later covered more in-depth by Spadaro *et al.*,[27] these defects originate from the reordering of vacant sites as growth proceeds along {101}-type facets, manifesting as a 120° crystal rotation. The apparent threefold symmetry of the {101} family of planes results in only a small energetic splitting between competing orientations, which can be overcome by kinetic growth processes, leading to the formation of rotated domains. This interpretation is consistent with density functional theory (DFT) calculations, which indicate a negligible energetic penalty for the formation of rotated (101) interfaces, as evidenced by the similar adhesion energies of rotated and non-rotated domain pairs.[27] However, alternative rotated interfaces that develop during domain coalescence are shown to introduce shallow electronic states at approximately 0.1 eV, 0.21 eV and 0.38 eV above the valence band, which are in good agreement with experimentally observed emission features reported in the literature. [28–35]

In this work, we study in detail an additional mechanism of Zn vacancy rearrangement occurring along the stacking direction of cation and anion layers (the c-axis), which manifests as planar defects within the $Zn_3P_2$ crystal. Planar defects have previously been reported in vapour–liquid–solid (VLS)-grown $Zn_3P_2$ nanowires and selective-area epitaxy (SAE) thin films.[36,37] Based primarily on their appearance as parallel contrast features in transmission electron microscopy (TEM) images, these defects were referred to as stacking faults. However, their exact crystallographic nature and influence on the structural and electronic properties of the material remained unexplored.

Only recently, Hagger *et al.* reported high densities of planar defects in metal–organic chemical vapour deposition (MOCVD)-grown $Zn_3P_2$ grains on graphene, which they identified as antiphase boundaries (APBs).[38] Through spatially resolved cathodoluminescence (CL) spectroscopy, they observed local variations in emission intensity and peak energy that correlated with the APB density. These results suggested that APBs perturb the local electronic structure, likely through the introduction of potential fluctuations that modulate sub-bandgap emission. However, it remained unclear whether the APBs themselves directly generate the observed electronic states through local bonding distortions and structural irregularities or instead act indirectly as preferential sites for the formation or segregation of optically active point defects. Establishing their exact role is therefore essential for understanding the defect physics of $Zn_3P_2$ and for identifying the mechanisms limiting its optoelectronic performance. Our study aims to bridge this gap by elucidating the atomic and electronic structure of these planar defects and their impact on the material properties through DFT calculations. This is particularly important in light of the detrimental effects that planar defects, including stacking faults and related extended defects, have been shown to exert in c-Si,[39] CdTe,[40,41] CIGS,[42] and metal–halide perovskite solar cells,[43–45] which represent the principal competing PV technologies to $Zn_3P_2$-based devices.

We begin by presenting a comprehensive structural characterization of planar defects in $Zn_3P_2$ using selected-area electron diffraction (SAED), aberration corrected high-angle annular dark-field scanning transmission electron microscopy (AC HAADF-STEM), and three-dimensional (3D) atomic modelling. Examination of a large set of $Zn_3P_2$ samples grown by molecular beam epitaxy (MBE) and MOCVD under different growth conditions and on different substrates reveals a high density of planar defects throughout the as-grown material, indicating that defect-free $Zn_3P_2$ is difficult to achieve using conventional growth approaches. To describe the modified atomic arrangement associated with planar defects, we develop a discrete structural model based on the pseudo-cubic approximation and interpret it within the framework of order–disorder (OD) theory.[47] Following this description, the bulk crystal corresponds to a maximum order polytype obtained by the periodic stacking of two distinct layer types.[47] In contrast, planar defects originate from a local disruption of this stacking sequence through the insertion of an additional layer that breaks the bulk long-range ordering.

First-principles calculations reveal an almost exact energetic degeneracy between the pristine bulk and faulted configuration, in excellent agreement with OD theory and the high density of planar defects observed experimentally. Additionally, this result establishes a direct connection between planar defects and rotated domains, showing that both types of extended defects can form at negligible energetic cost through local reordering of the vacant sites in the Zn sublattice during growth.

Further DFT calculations reveal an almost identical optoelectronic response between the pristine and faulted systems. Electrostatic potential profiles across the defect plane show no significant local perturbation, indicating that intrinsic planar defects are not directly responsible for the CL intensity variations reported by Hagger *et al.*[38] Future studies will be directed in assessing the indirect impact of extended planar defects as preferential segregation sites for point defects. These findings go beyond the specific system studied here: They highlight the importance of evaluating the indirect impact of apparently harmless structural defects on the optoelectronic properties of semiconductors.

# 2. Methods

## 2.1 Electron microscopy

Electron transparent lamellae of VLS nanowires, SAE nanopyramids, SAE thin films and $Zn_3P_2$ crystals grown on graphene by Van der Walls epitaxy were obtained via focused ion beam (FIB) processing using a Helios UX 5 FIB system. More information on the initial synthesis of the samples can be found in refs.[28,30,48,49] A pre-deposition of 0.5 µm of carbon (nominal thickness) and 0.5 µm of tungsten(W) (nominal thickness) was performed with the electron gun inside the FIB system prior to the lamella processing. This procedure aimed at limiting the contamination from the thick ion-deposited W protective layer during the final stages of the thinning. Additionally, the contrast given by the carbon when imaging via SEM ICE detector, during the final stages of the thinning, helped to locate the nanostructures. Selected area electron diffraction (SAED) analysis was performed in the transmission electron microscope (TEM) Tecnai F20 operated at 200 kV. A selected area aperture with a physical diameter of 10 µm was used to isolate regions of interest at a magnification of 6300 SA (this results in a probed area of approximately 30 nm in diameter). We find a rotational angle of 90° between the diffraction plane and the image plane. HAADF AC -STEM and iDPC-STEM images were obtained using a Thermo Fisher Scientific Spectra 300 double corrected scanning transmission electron microscope (STEM) operating at 300 kV. 3D atomic models were created by using the on-line Rhodius software.[50,51]

## 2.2 Ab initio and machine-learning accelerated calculations

All first-principles calculations were performed using density functional theory as implemented in the Vienna Ab initio Simulation Package (VASP).[52–54] We employed the PBEsol exchange-correlation functional.[55,56] The plane-wave basis set was expanded up to a kinetic energy cutoff of 350 eV. Brillouin zone integration was handled using a Gamma-centered 4 × 4 × 1 *k*-point mesh. The Zn potential treated 12 valence electrons ($3d^{10}4s^2$) as explicit valence states, while the P potential accounted for 5 valence electrons ($3s^23p^3$). We conducted the calculations with the inclusion of spin-orbit coupling.

To supplement the first-principles calculations, we utilized the machine-learned interatomic potential (MLIP) known as eSEN.[57] This model corresponds to the checkpoint published by Meta under the name esen_30m_oam.pt. The MLIP allowed for efficient initial screenings of various configurations before performing high-accuracy validation with DFT.

The starting point for this study was the $Zn_3P_2$ unit cell obtained from the Materials Project database under entry mp-2071.[58] We fully optimized the cell geometry and internal ionic degrees of freedom using DFT to ensure a consistent reference for the bulk material.

Supercells were constructed by extending the optimized unit cell along the ***c***-direction while keeping the ***a***- and ***b***-basis vectors constant. This construction restricts the lateral symmetry of the (001) planes. The phosphorus sublattice was held fixed in its ideal positions. In contrast, we varied the positions of the zinc atoms within the potential sublattice sites defined by

$$(x, y) = [(0, \tfrac{1}{4}), (0, \tfrac{3}{4}), (\tfrac{1}{4}, 0), (\tfrac{1}{4}, \tfrac{1}{2}), (\tfrac{1}{2}, \tfrac{1}{4}), (\tfrac{1}{2}, \tfrac{3}{4}), (\tfrac{3}{4}, 0), (\tfrac{3}{4}, \tfrac{1}{2})].$$

We made use of the pymatgen Python package[59] for the construction of these models and for the subsequent symmetry analysis of the stacking faults.

The internal ionic degrees of freedom for all defect models were relaxed until the forces on each atom were smaller than 1 meV/Å. We did not relax the cell geometry during these steps to respect the lateral constraints of the bulk crystal. Preliminary tests showed that optimizing the c-basis vector to allow for relaxation normal to the stacking fault resulted in negligible energy changes. Consequently, we fixed the cell dimensions for all reported calculations for computational efficiency.

We determined the formation energies of various stacking defects using an extrapolation scheme similar to standard surface energy calculations.[60–62] This approach ensures that the results are independent of the supercell dimensions or the distance between periodic images of the faults. We used the following linear relationship:

$$E_{SF} = 2 \cdot A \cdot E_f + N_{SF} \cdot e_{bulk}$$

Here, $E_{SF}$ represents the total energy of a supercell containing $N_{SF}$ atoms with a specific stacking fault. $E_f$ is the formation energy per unit area, $A$ is the cross-sectional area, and $e_{bulk}$ is the energy per atom of the pristine crystal. By calculating $E_{SF}$ for supercells of varying sizes, we extracted $E_f$ through linear extrapolation. This method provides higher numerical stability than a direct comparison with a separate bulk calculation.

The densities of states (DOS) were computed using the tetrahedron method with Blöchl corrections without smearing, as implemented in VASP. We utilized a significantly finer $k$-point mesh than the one used for ionic relaxations. For the pristine bulk, we used a mesh of $8 \times 8 \times 5$. The DOS for the stacking fault as found in experiment was calculated using a 120-atom supercell and a $k$-point mesh of $8 \times 8 \times 2$.

The local electrostatic potential was calculated by taking the plane average of the potential along the $z$-coordinate. We then applied a macroscopic averaging technique to identify any potential shifts or dipoles induced by the structural disruptions at the fault interface.

While the main text focuses on the stacking fault observed in experimental data, we have included a comprehensive analysis of other potential stacking configurations in the Supplementary Information.

# 3. Results and discussion

## 3.1. (S)TEM characterization of $Zn_3P_2$ planar defects

In this section, we investigate the structural properties of the observed planar defects using (S)TEM analysis, taking as a representative example a $Zn_3P_2$ thin film grown by MOCVD-SAE on an InP (100) substrate. Figure 1a presents a SEM side-view image of the sample, along with a large field-of-view HAADF STEM image of a FIB prepared lamella (see Methods), shown in the inset. Selected area electron diffraction (SAED) patterns acquired from different regions of the sample reveal two predominant crystallographic orientations, indexed as the (001)×[010] and (1-10)×[111] zone axes. These orientations correspond to the rotated domains in $Zn_3P_2$.[31] Figures 1b and 1c show the experimental and simulated diffraction patterns from the two domains. The diffraction patterns of the two orientations are identical apart from the presence of additional reflections that appear exclusively when the crystal is oriented along the [010] zone axis. In addition, the diffraction spots when observing along the [010] zone axis exhibit pronounced diffuse streaking along the *c*-axis. Such diffuse intensity indicates the presence of planar defects oriented perpendicular to the streaking direction and distributed with zero ordering across the probed area.[49–51]

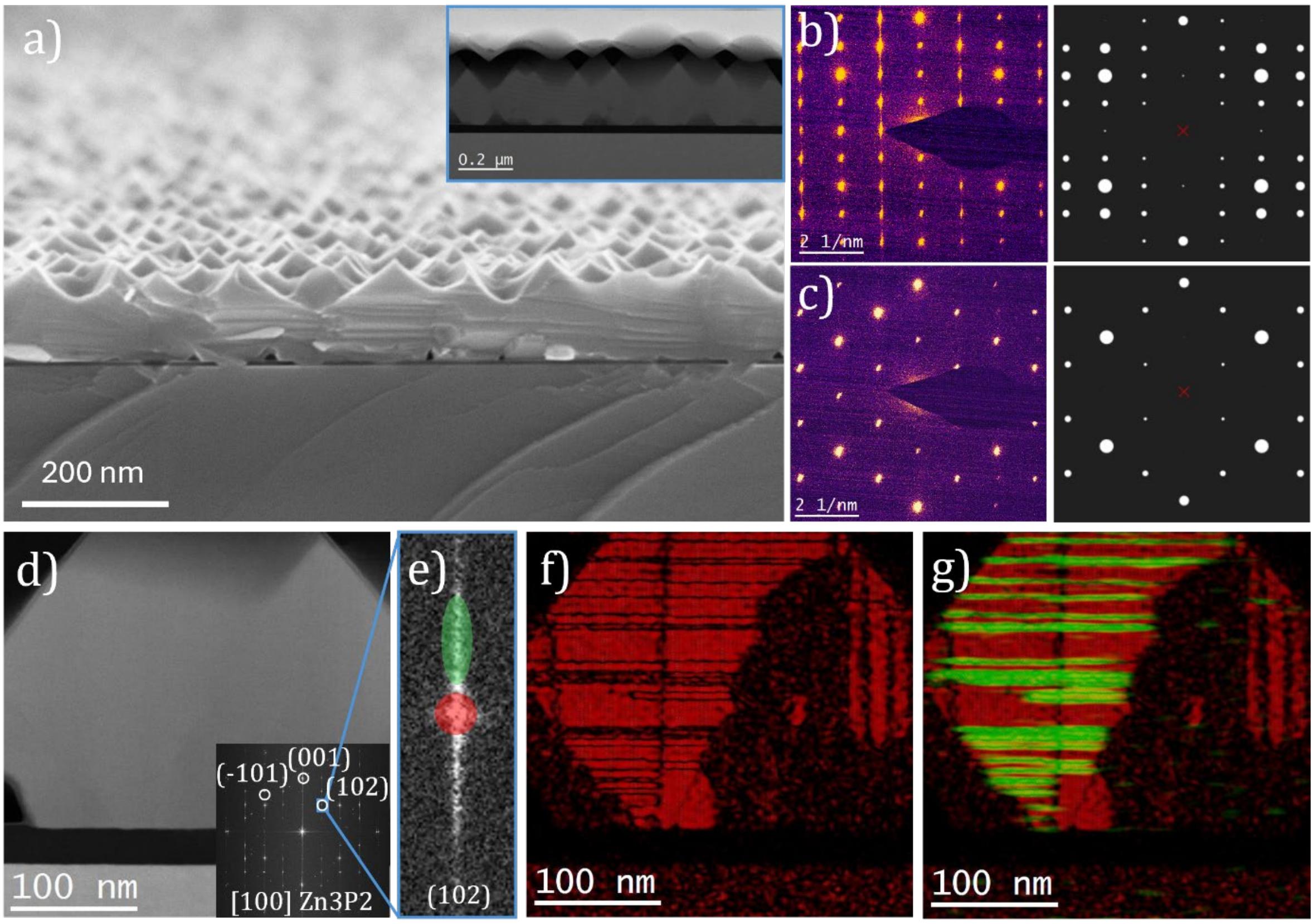

**FIGURE 1:** (a) Side-view SEM image of a SAE $Zn_3P_2$ thin film grown by MOCVD. In the inset, the HAADF image of the FIB cross-section of the thin film is shown. b,c) Experimental and simulated SAED patterns of a region of the crystal oriented along the <100> and <111> zone axis, respectively. Owing to the finite lamella thickness, plural scattering gives rise to nominally forbidden reflections in the experimental diffraction patterns for both orientations (d) HAADF AC-STEM image of a region of the thin film with the corresponding FFT in the inset. A detail of the reflection relative to the (102) plane is shown in (e), from which the Bragg filtered color maps in (f) and (g) are obtained.

Generally, the nature of planar defects can be understood by following the extinction rule for the diffuse streaks.[66] Intrinsic defects generate streaking only for specific Bragg reflections and can be described by a displacement vector $\boldsymbol{R}$ lying within the fault plane.[67] $\boldsymbol{R}$ obeys the extinction criteria, which state that when $\boldsymbol{g} \cdot \boldsymbol{R}$ = integer, the streaks are extinguished, whereas when $\boldsymbol{g} \cdot \boldsymbol{R} \neq$ integer, the streaks are present, where $\boldsymbol{g}$ denotes the diffraction vector. In our case, streaking is observed for reflections of the form ($h$0l) with $h$ odd. This condition constrains the displacement vector to have a half-integer component along [100], such that $\boldsymbol{R} = [½, v, 0]$, where $v$ remains undetermined (the vector $\boldsymbol{R}$ is expressed in fractional coordinates relative to the lattice parameters $a = b$). The latter reflects the fact that, for the [010] zone axis, the extinction condition is insensitive to the component of $\boldsymbol{R}$ along [010]. Since $Zn_3P_2$ exhibits tetragonal symmetry, where [100] and [010] are equivalent, the displacement vector can also contain a half-integer component along [010]. Consequently, the displacement vector is equally consistent with $\boldsymbol{R} = [½, 0, 0]$ and $\boldsymbol{R} = [½, ½, 0]$. It is important to underline that local planar defects can break the 4-fold symmetry of the crystal: while the lattice symmetry allows $\boldsymbol{R} = [½, 0, 0]$ and $\boldsymbol{R} = [0, ½, 0]$ as equivalent variants, it does not imply that an individual defect must be described by $\boldsymbol{R} = [½, ½, 0]$.

To visually illustrate the presence of planar defects throughout the crystal, we obtained Inverse Fast Fourier Transform (IFFT) images starting from the atomic resolution HAADF STEM image in Figure 1d. The reconstructed images obtained by masking Bragg reflections with *h* odd in the corresponding FFT, selectively enhance the contrast of planar defects, as they introduce a phase discontinuity for the corresponding Fourier components in the frequency space. For instance, when considering the (102) reflection (enlarged for clarity in Figure 1e), masking the Bragg spot produces horizontal dark stripes in the domain oriented along the [010] zone axis (Figure 1f). In contrast, the associated diffuse intensities, arising from the relaxation of the Bragg condition at the fault planes, map onto these regions in a complementary manner (Figure 1g). Note that the domains oriented along the [111] zone axis remain dark, since the (102) reflection is not excited in this orientation.

Overall, we find that planar defects are distributed throughout the structure, with no indication of short- or long-range ordering, and are laterally terminated by rotated domains and/or crystal edges. Additionally, to first approximation based on the extinction criteria, planar defects in $Zn_3P_2$ can be described by simple displacement vectors lying within the (001) plane. These findings are consistent across all $Zn_3P_2$ samples examined in this work, regardless of the growth approach (SAE or VLS via MBE or MOCVD; see Figure SI1), growth conditions (temperature and precursor partial pressures), and substrate material (InP, Silicon, and Graphene; see Figure SI2). From this systematic characterization, we conclude that the formation of planar defects is an intrinsic feature of conventionally grown $Zn_3P_2$, likewise the case of rotated domains,[27] making the realization of structurally defect-free material unlikely via conventional growth approaches.

To investigate the crystallographic nature of planar defects in $Zn_3P_2$ and verify their description in terms of the proposed displacement vectors, we acquired atomic-resolution HAADF AC-STEM images of the crystal oriented along multiple zone axes (Figure 2). Owing to the incoherent nature of high-angle electron scattering, this imaging mode provides a direct correlation between image intensity and projected atomic column density, with a strong dependence on the atomic number Z.[68] This characteristic enables direct visualization of the Zn atomic columns in $Zn_3P_2$. Focusing on the [100] orientation shown in Figure 2a, the fault plane is identified at the position of contrast discontinuity in the reconstructed IFFT image (Figure 2b). The atomic resolution HAADF STEM image in Figure 2c reveals that the Zn atomic columns above the fault plane (indicated by the dashed white line) are shifted by a half-translation along the [010] direction (purple arrow). The Zn atoms arrangement from two adjacent displaced unit cells are highlighted with coloured circles for clarity (Figure 2c). The 3D atomic model in Figure 2d is then built based on the displacement vector $\boldsymbol{R} = [0, ½, 0]$. In the Supporting Information, we verify the validity of the atomic model for both Zn and P atoms by superimposing it to the STEM integrated-differential phase contrast (i-DPC) image of the crystal oriented along the [100] zone axis (Figure SI3). The 3D atomic model accurately reproduces the atomic arrangements observed along the [120] and [110] zone axes (Figure 2), obtained by 26.57° and 45° rotations about the c-axis, respectively. Additional zone-axis projections consistent with the model are shown in Figure SI4.

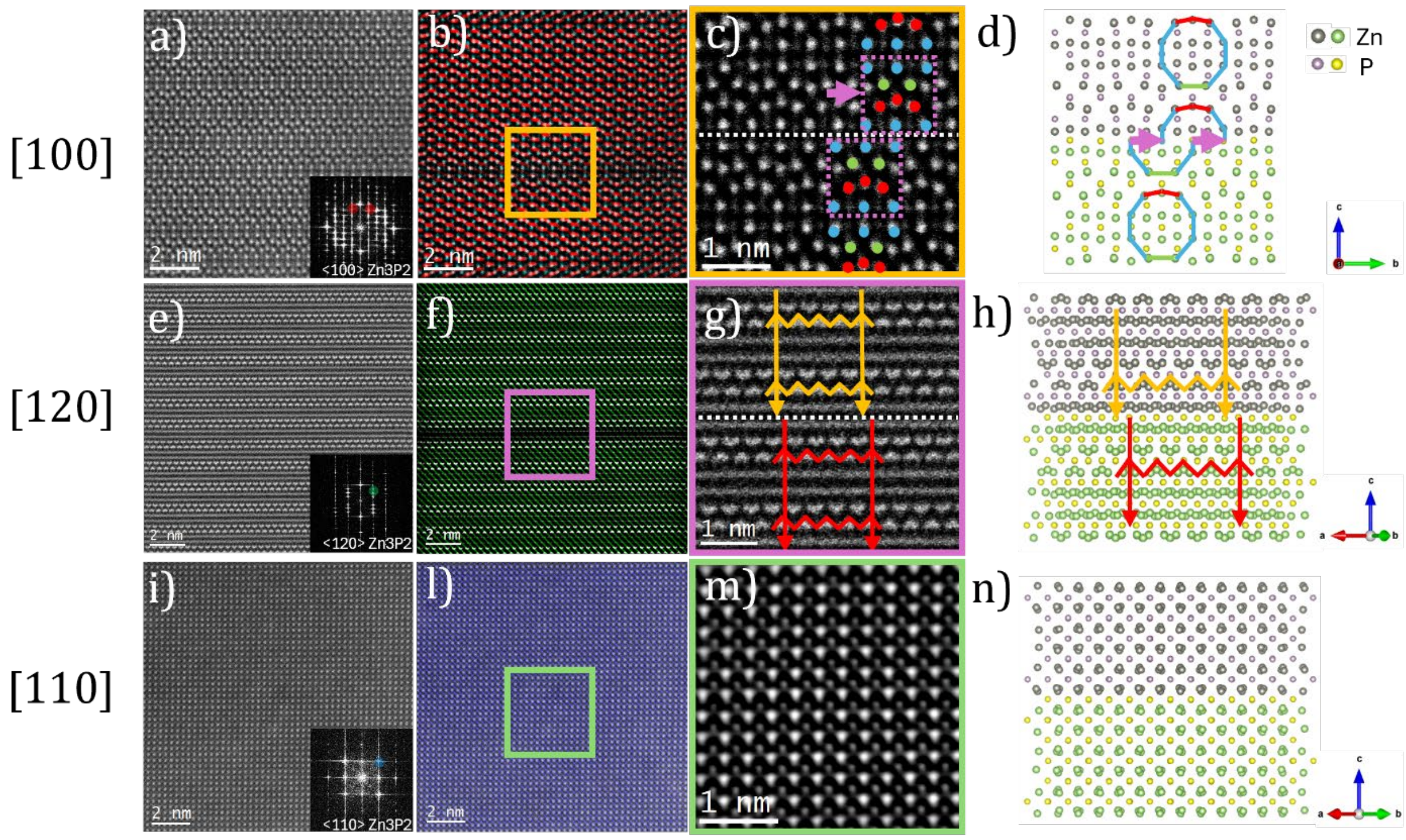

**FIGURE 2:** (a) Atomic resolution AC HAADF-STEM images of the $Zn_3P_2$ crystal oriented along the <100> zone axis with the FFT in the inset. (b) Corresponding filtered map obtained by masking (*hkl*) reflections that show streaking. (c) Detailed AC HAADF-STEM image highlighting the faulted region and fault plane with white dashed line. (d) 3D atomic model of the faulted region, which highlights with different colours the atoms from the two displaced regions. The same sequence of images is repeated for the <120> (e-h) and <110> (i-n) zone axis.

As a first approximation, the extinction criteria show that intrinsic planar defects in $Zn_3P_2$ are equally consistent with displacement vectors of type R = [0, ½, 0] and R = [½, ½, 0]. However, depending on the reference position of the fault plane, a single defect may not be described by R = [½, ½, 0]. In Figure 3, we show the pristine and defective structures together with the corresponding 3D atomic models. By applying both displacement vectors to the atoms above the fault plane indicated by the blue dashed line (same reference plane used in Figure 2), we find that R = [½, ½, 0] disrupts the chemical ordering and generates energetically unfavourable nearest-neighbour configurations, whereas R = [0, ½, 0] preserves the local atomic coordination. Interestingly, if the fault plane is shifted by ±1 Zn atomic layer (red dashed lines in Figure 3), both R = [0, ½, 0] and R = [½, ½, 0] preserve the local chemical ordering across the boundary. 3D atomic models illustrating the effect of each displacement vector applied to the pristine bulk are provided in the Supporting Information (Figure SI5). This dependence on the fault plane reference position stems from the non-symmorphic screw axis operation that characterises the $P4_2/nmc$ space group of $Zn_3P_2$, in which each 90° rotation is intrinsically coupled with a fractional translation of **c**/2.[69]

Based on this analysis, we find that the crystallographic description of the planar defect is not unique. Although the diffuse-scattering extinction criteria constrain the relative translation across the interface, the displacement vector depends on the choice of reference plane used to define the boundary. Consequently, diffraction contrast and projected HAADF-STEM imaging alone cannot uniquely determine the three-dimensional interfacial geometry across the fault plane. Nonetheless, the fact that the defect is not visible along zone axes that do not project the Zn vacant sites (see [110] orientation in Figure 2, and [111] orientation in Figure SI4) suggests that planar defects are best interpreted as a rearrangement of vacant sites within the Zn sublayers stacked along the c-axis. In the following sections, we propose two equivalent models of the three-dimensional bulk structure of $Zn_3P_2$, which account for altered stacking sequences.

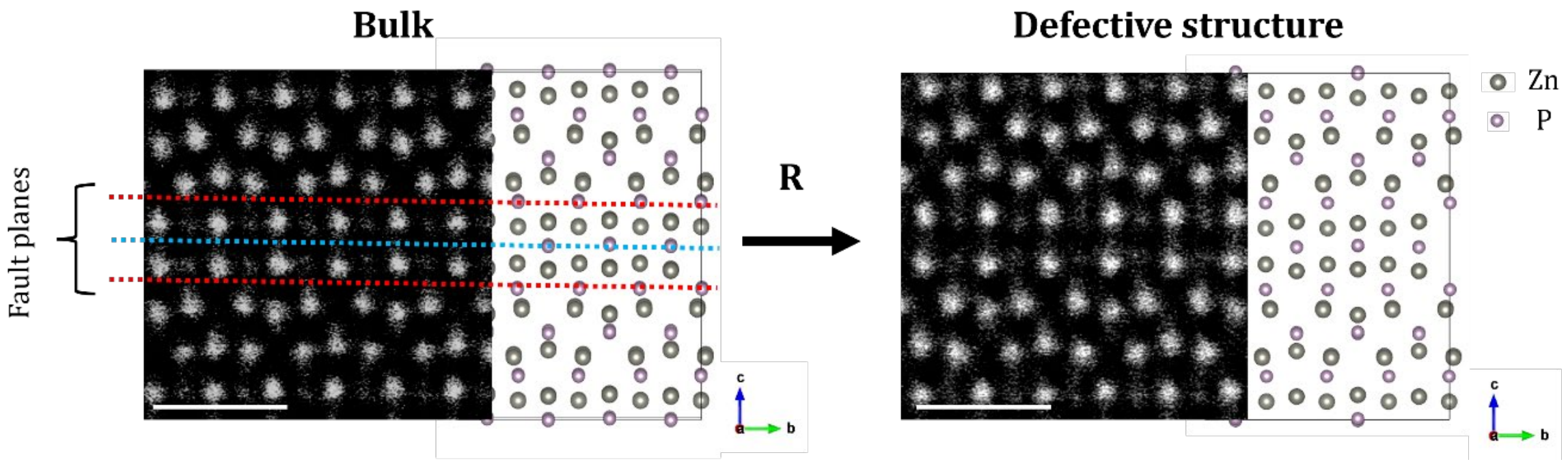

**FIGURE 3:** Atomic-resolution HAADF-STEM images of $Zn_3P_2$ in the pristine and defective configurations, viewed along the (001)×[100] zone axis, together with the corresponding 3D atomic models. The reference fault planes are indicated by coloured dashed lines in the pristine structure. The defective structure is obtained by displacing the atoms above the blue fault plane by **R** = [0, ½, 0]. Equivalently, it can be obtained by displacing the atoms above the red fault planes by either **R** = [0, ½, 0] or **R** = [½, ½, 0]. Scale bar of the HAADF-STEM images is 1 nm.

## 3.2. Polytypism in $Zn_3P_2$

To elucidate the nature of the observed planar defects, we propose a model based on the extended elementary cell (EEC) of $Zn_3P_2$, which is commonly used to describe its behavior during epitaxial growth (Figure 3).[7,8,10,28,70,71] The cell parameters of the EEC, (**a**,**b**,**c**), are related to those of the primitive cell, ($\mathbf{a_0}$,$\mathbf{b_0}$,$\mathbf{c_0}$), as: $\mathbf{a} = \mathbf{a_0} - \mathbf{b_0}$, $\mathbf{b} = \mathbf{a_0} + \mathbf{b_0}$ and $\mathbf{c} = \mathbf{c_0}$; where we assume that a = $c_0$ and atoms occupy their ideal positions, thus resulting in a cubic unit cell with size $\sqrt{2}a_0 \times \sqrt{2}a_0 \times \sqrt{2}a_0$. In the coordinate system of the EEC, the crystal structure can be described as a sequence of two alternating packets of atomic layers.[23,47,72–75] Each packet consists of two close-packed anion layers interleaved with two cation layers. These packets can be represented in two equivalent ways. In the first description, they correspond to cubic fragments of zinc blende–type (50% Zn occupancy) and fluorite-type (100% Zn occupancy) structures arranged in a checkerboard pattern (denoted as packets X and Y). Alternatively, they can be described as cubic fragments of a defect fluorite structure with 75% Zn occupancy, also arranged in a checkerboard pattern (denoted as packets W and Z). Although X (or W) and Y (or Z) are not equivalent, they are related by partial translations of type ½(**a**+**c**) and ½(**b**+**c**). Accordingly, bulk $Zn_3P_2$ can be equivalently described by the stacking sequences: …/XY/XY/… or …/WZ/WZ/… (Bulk structures in figure 3). Within each packet,

Zn vacant sites are located at diagonally opposite corners of the faces of the constituent cubic fragments. These vacant sites alternate between the two diagonal orientations in successive packets

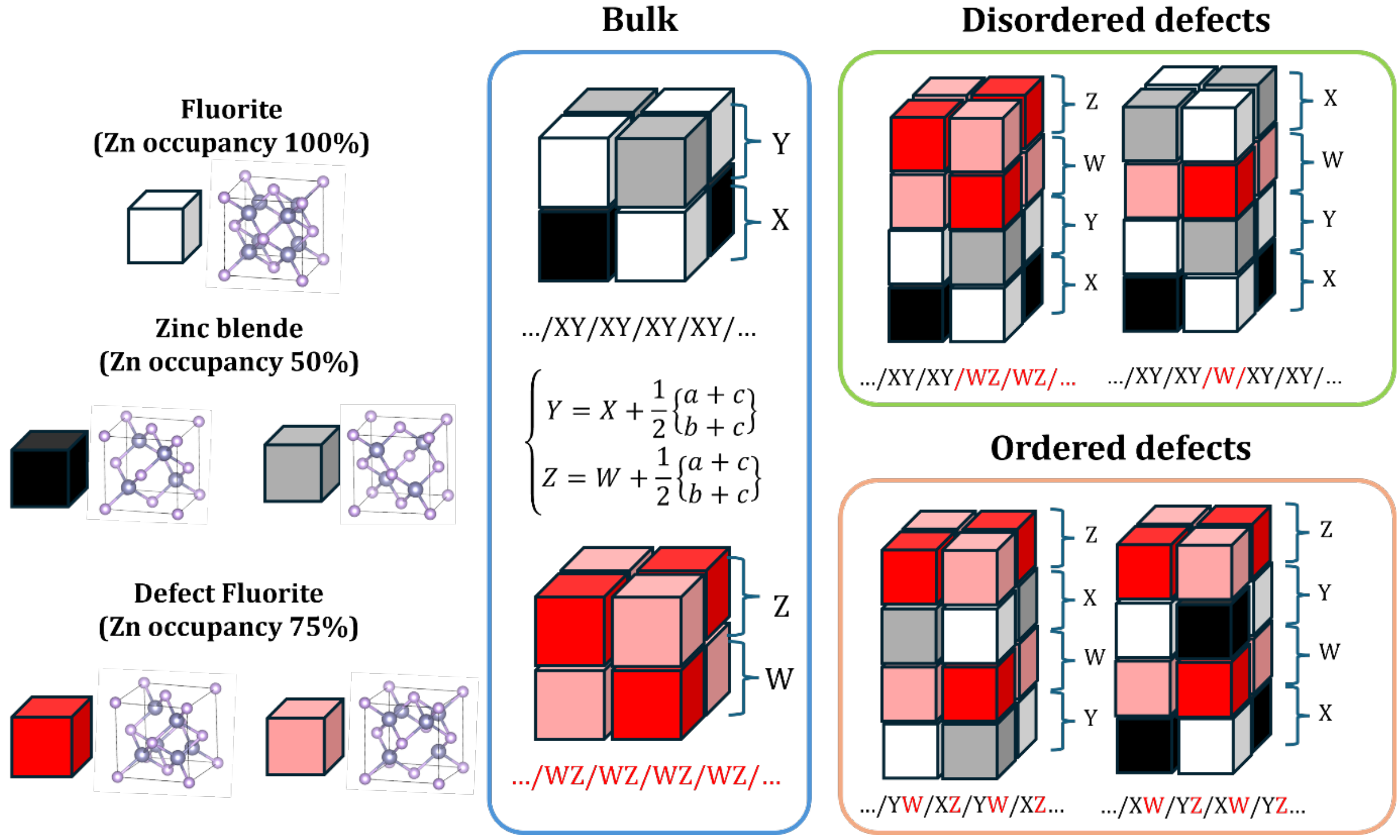

**FIGURE 4:** Schematic model of bulk and faulted $Zn_3P_2$ based on the stacking of non-equivalent packets. Each packet is built by the periodic arrangement of constituent building blocks, i.e., fluorite, zinc blende and defect fluorite unit cells.

along the c-axis.

We now consider the bulk structure described by the alternating X and Y packets. At the fault plane, the insertion of a packet of type W disrupts the periodic sequence. Beyond the fault, the bulk structure may be restored either by returning to the X–Y sequence or by continuing with the W–Z sequence (disordered defects in Figure 3). Following this construction, the formation of fully ordered structures (ordered defects in Figure 3), can be proposed as arising from the periodic mixing of packets of different types. Representative examples include the sequences …/YW/XZ/… and …/XW/YZ/… (Figure 3). As discussed above, inspection of the SAED patterns reveals no evidence of short- or long-range ordering in the probed areas. This is supported by the presence of continuous diffuse streaking of the Bragg reflections corresponding to (hkl) with h odd. The presence of discrete satellite reflections symmetric about the Bragg positions would instead be indicative of short-range ordering.[76] Nonetheless, ordered stacking sequences are locally observed (Figure SI6). In the Supporting Information, we simulate the SAED patterns of superlattice structures obtained by the periodic ordering of faulted sequences. Since the phase relationship is enforced coherently across the entire crystal, destructive interference completely cancels some bulk reflections (forbidden Bragg reflections), while the new larger periodicity simultaneously generates additional satellite reflections at fractional positions symmetric to the forbidden Bragg reflections along the direction perpendicular to the fault plane (Figure SI7).

Our experimental observations can be interpreted within the framework of the order–disorder (OD) theory.[77–80] According to the OD theory, polytypic structures can be described as assemblies of structural parts that are periodic in two dimensions, commonly referred to as OD packets.[46,81] These packets can be stacked in multiple translationally non-equivalent ways while preserving the nearest-neighbour relationships. The packets we defined for $Zn_3P_2$ satisfy this formalism as they are transformed into one another by the pseudo-translations in Figure 3 and their stacking preserves the coordination of P atoms with Zn atoms (and vice versa). OD theory also states that structures with the maximum degree of order (MDO) have stacking sequences that involve the minimum number of symmetrically non-equivalent packets. For $Zn_3P_2$, this condition is satisfied by the bulk structures in Figure 3. The corresponding diffraction patterns exhibit sharp Bragg reflections (Figure SI7), indicative of long-range periodic order. In contrast, non-MDO polytypes arise when the stacking sequence includes more distinct packets than the minimum geometrically required. This alteration can either disrupt the periodic ordering of the bulk structures or generate periodic superstructures that possess short- or long-range ordering. The former case is exemplified by the "disordered defects" $Zn_3P_2$ structures in Figure 3, which are consistent with our experimental observations. Because the altered stacking sequence is non-periodic (i.e., its repeat period is effectively infinite), a subset of Bragg reflections exhibits diffuse streaking rather than remaining sharp. Alternatively, the altered sequence may introduce superstructure-like ordering, as in the case of the "ordered defects" structures in Figure 3, resulting in extra sharp reflections in the diffraction pattern (see Figure SI7). This situation is never observed experimentally.

OD theory further predicts that MDO and non-MDO structures have comparable formation energies, since the local bonding environments are preserved in both cases.[80] Nevertheless, MDO structures are generally observed more frequently. While, in a first approximation, only the interactions within two atomic layers need to be considered, in practice, longer-range interactions also play a role. As a result, structures containing the minimum number of distinct packets are energetically favoured, due to their higher geometric symmetry. However, during crystal growth, kinematic effects can influence stacking sequences in an uncontrolled way, sometimes promoting the formation of disordered, non-MDO structures, as observed in $Zn_3P_2$. A similar argument can be made concerning rotated domains. Within the pseudo-cubic framework, rotated domains emerge naturally due to the apparent three-fold rotational symmetry around the direction perpendicular to the {101} planes. The almost exact degeneracy is lifted by the presence of ordered Zn vacant sites in the crystal structure. As a result, the formation of rotated domains is governed by kinetically driven processes that enable overcoming the small energetic difference between orientations in a stochastic way during growth.

## 3.3 Energetics and electronic structure from ab initio calculations

The description of $Zn_3P_2$ in terms of fluorite-type cubic sub-units is effective for microscopic analysis. It simplifies the interpretation of Bragg reflections by allowing direct comparison with known structures. However, for ab initio simulations, we utilize an equivalent description based on (001) layers. This representation provides an alternative description of the observed stacking fault and simplifies the construction and description of the employed supercell models.

The conventional unit cell of $Zn_3P_2$ consists of eight distinct layers stacked parallel to the (001) plane. These comprise four phosphorus layers and four zinc layers, arranged in an alternating sequence. The four Zn sublayers differ based on the arrangement of Zn vacancies relative to the fluorite parent structure. We label these Zn layers as A, B, C, and D, as illustrated in Figure 4. In this framework, a stacking fault is defined as a deviation from the conventional Zn sublayer ordering.

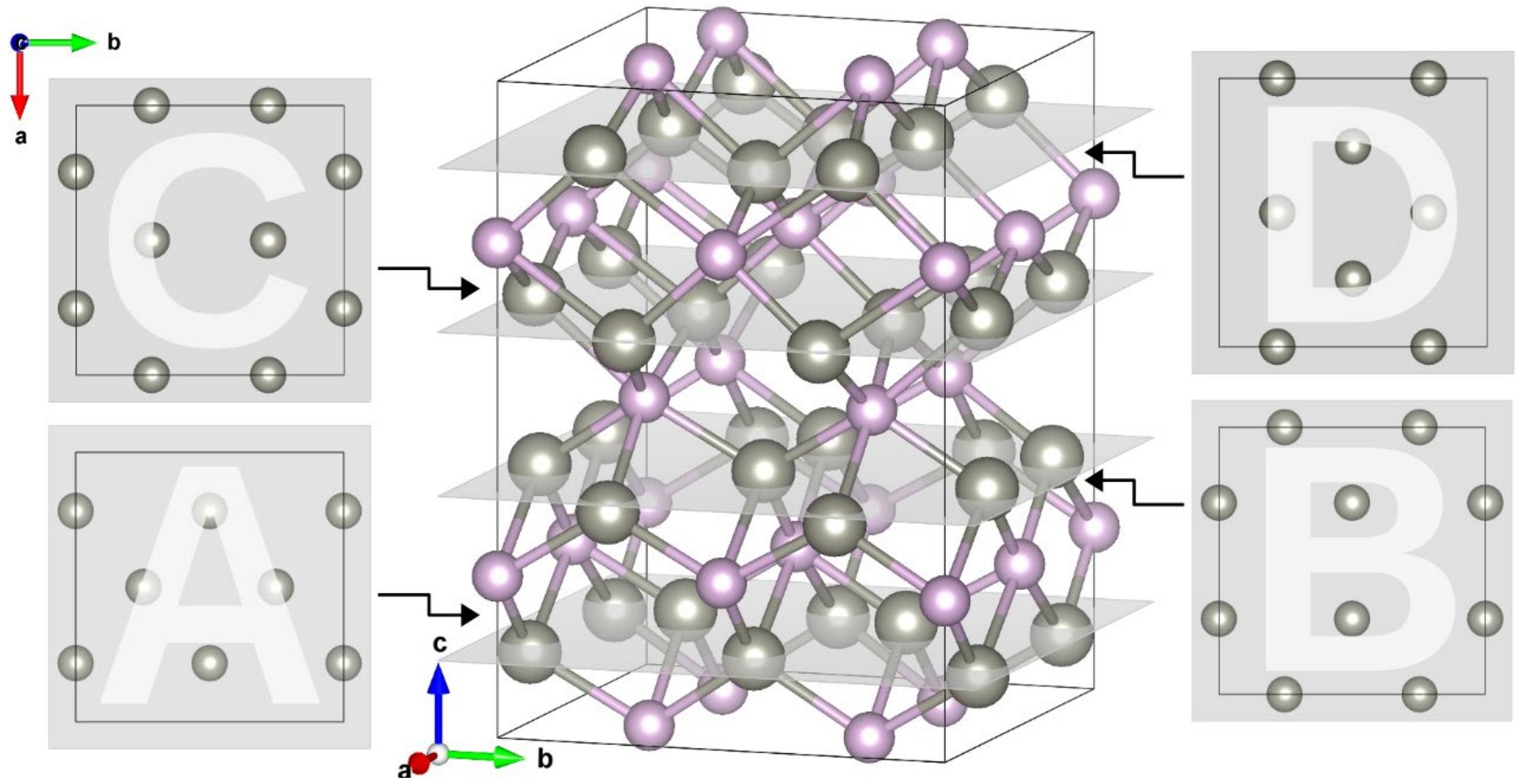

**FIGURE 5**: Labelling convention for (001) Zn sublayers in the $Zn_3P_2$ unit cell. Sublayers A to D are labelled in order of increasing *z*-coordinate within the conventional unit cell.

In isolation, the four sublayers A, B, C, and D are symmetry-equivalent and can be mapped onto one another through specific transformations: Sublayers A and D, as well as B and C, are related by a 90° rotation about the *z*-axis. Furthermore, the pairs A, B and C, D are related by translations of $\boldsymbol{a}/2$ and $\boldsymbol{b}/2$, respectively. This layer-based description is consistent with the previously discussed cubic sub-unit model. The symmetry operations relating these four layers account for the ambiguity in defining the conventional structure as a combination of X and Y packages (fluorite and zinc blende) rather than W and Z packages (defective fluorite).

We performed an exhaustive combinatorial exploration to identify various valid stacking fault candidates. We assumed that P atoms maintain a six-fold Zn coordination to remain consistent with the conventional structure. This condition greatly reduces the space of potential configurations, and at the same time suppresses planar defects associated with dangling bonds, which are unlikely to be stable. Combined with the lateral periodicity of the unit cell, only eight possible stacking faults fulfil this condition. Detailed descriptions of all eight models are provided in the Supplementary Information.

By further requiring that only Zn sublayers of types A through D are present, we can further narrow it down to only three possible stacking fault models. One of these three configurations corresponds exactly to the stacking fault observed in the experimental STEM data and is characterized by a non-sequential layer transition from A to B to D (instead of A to B to C), or C to D to B (instead of C to D to A), followed by a subsequent reversal of the stacking order. Due to the reversal in the stacking sequence, any supercell used as a model must contain at least two such stacking faults to satisfy periodic boundary conditions in the [001] direction.

The calculated formation energy for the observed stacking fault is exceptionally low at only 2.(5) mJ/m². While not common, formation energies this low have been previously reported for SiC.[82,83] Such low formation energy implies that the chemical environment near the stacking fault is very similar to that in the pristine bulk. In contrast, various other candidate stacking faults investigated in this study exhibited formation energies exceeding 150 mJ/m². Given this very low energy, these stacking faults are expected to be abundant in grown $Zn_3P_2$ samples, a result in excellent agreement with experimental observations.

Having established a structural model of the stacking fault, we now examine its impact on the electronic structure. Specifically, we analysed whether the defect introduces new states into the band gap or affects carrier transport. Such features could significantly influence the performance of the material in photovoltaic applications.

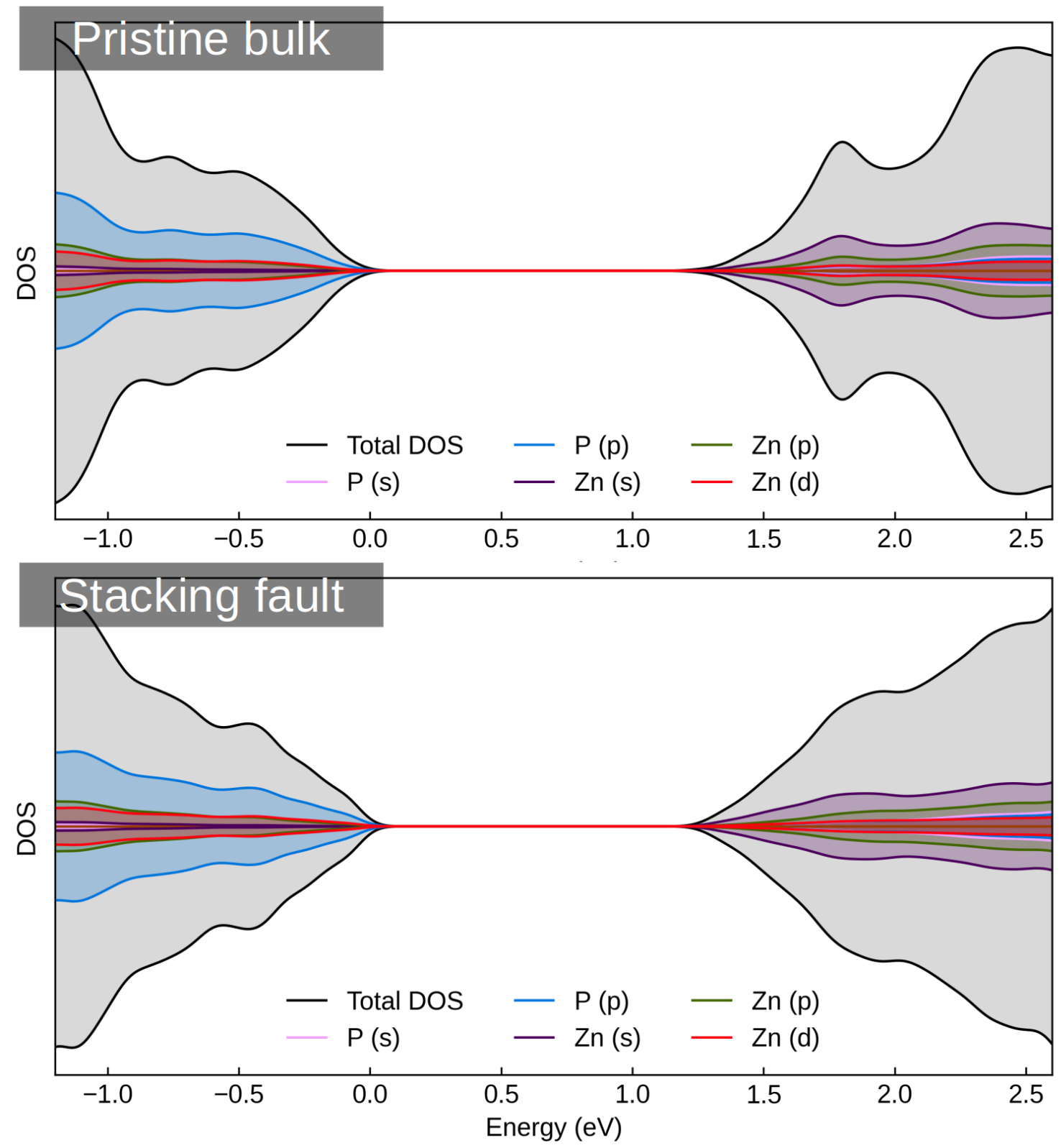

**FIGURE 6:** The total density of states (DOS) for the ideal Zn3P2 crystal (top) is compared to a supercell model incorporating two of the observed stacking faults (bottom). To remove sampling artifacts, a post-processing Gaussian broadening was applied with a standard deviation of 0.05 eV.

The conservation of the band gap and the absence of mid-gap states indicate that the fundamental electronic properties are robust against the described stacking disorder.

Figure 5 shows the computed density of states (DOS) for the pristine bulk phase compared to a supercell containing 120 atoms and two stacking faults separated by about 16.9 Å. This comparison shows that no new electronic states appear within the band gap. Additionally, the band gap magnitude remains virtually unchanged. This result is significant because the two other candidate stacking faults caused a substantial reduction in the band gap by introducing new states below the conduction band minimum (see Figure SI8). This suggest that the stacking fault observed in real samples is electrically benign and does not appear to contribute to unwanted carrier recombination despite its high abundance.

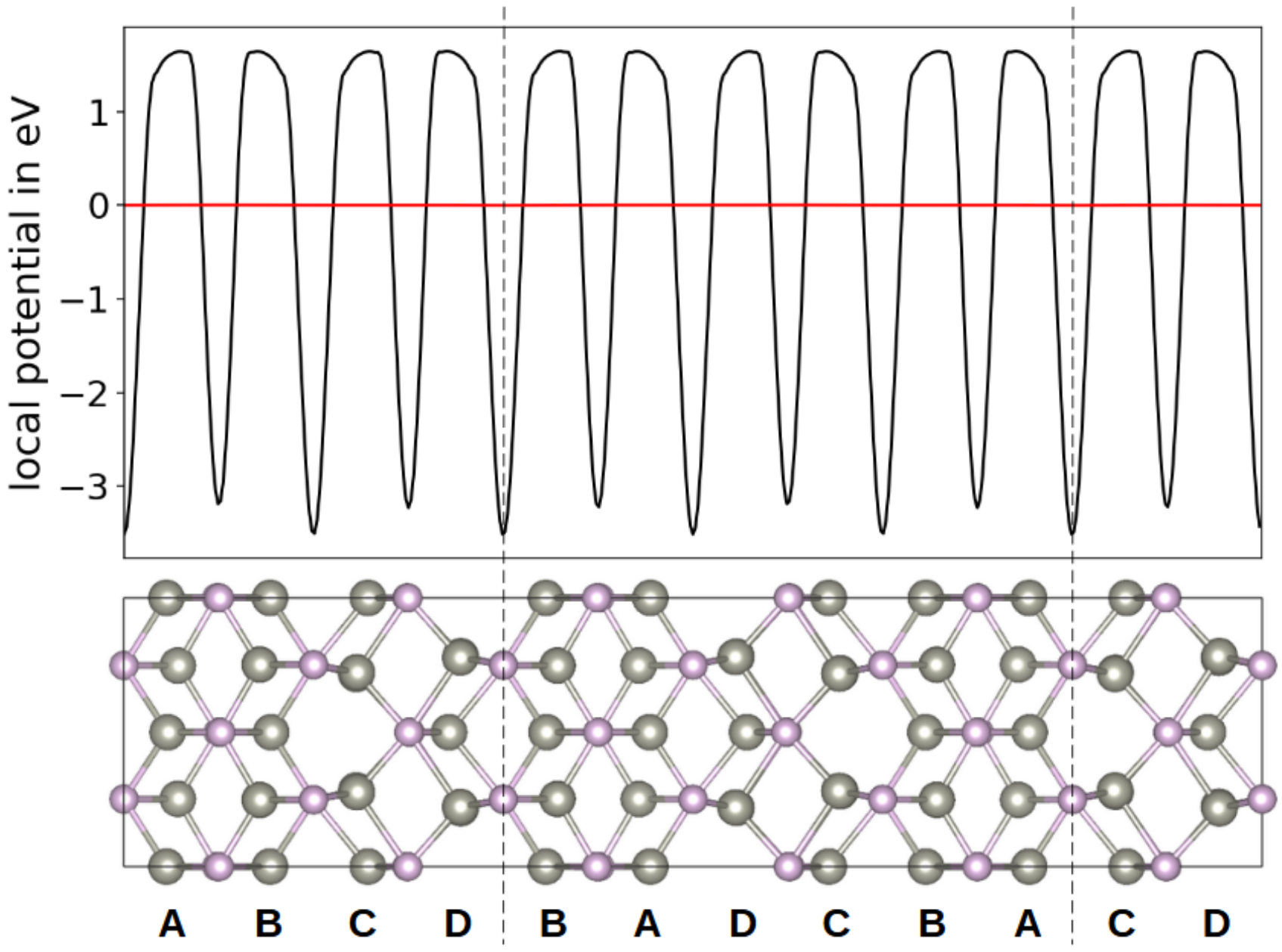

**FIGURE 7:** Plane-averaged electrostatic potential across $Zn_3P_2$ stacking faults. (Top) Local average potential (black line) and macroscopic average potential (red line) plotted along the *z*-coordinate. (Bottom) Corresponding atomic model of the supercell, aligned to the spatial scale of the potential plot, with (001) Zn sublayers labelled A to D. Vertical dashed lines denote the positions of the two stacking faults, where the non-conventional stacking sequence is introduced. The continuity of the macroscopic average across the boundaries indicates that the stacking faults do not introduce significant electrostatic disruptions or dipoles.

We further computed the layer-averaged local potential along the [001] direction to evaluate the impact of the fault on electronic transport. The results are presented in Figure 6. No significant potential barriers are observed at the defect site. This implies that these stacking faults do not impede carrier transport in the direction normal to the fault plane. Moreover, the absence of a discernible potential step suggests that the stacking fault is effectively nonpolar, in contrast to polarity-inverting planar defects that can generate electrostatic discontinuities and suppress conductance.[84]

Our findings demonstrate therefore that, despite the large occurrence of the investigated defect, it does not alone affect the electronic properties of the material, demonstrating their presence is not detrimental for PV applications. Further simulations are currently ongoing to investigate if the stacking faults here introduced and described could act as trap for other type of defects, such as point defects.

# 4. Conclusion

This study provides a comprehensive structural and electronic characterization of stacking faults in $Zn_3P_2$. By integrating atomic resolution HAADF-STEM imaging with ab initio density functional theory, we have identified the ABDC stacking sequence as the dominant planar defect. The prevalence of this specific configuration and its abundance in grown samples is elucidated through order–disorder theory and further supported by our energetic calculations, which reveal an exceptionally low formation energy of 2–3 $mJ/m^2$. Explicit computation of alternative stacking sequences confirms that the ABDC fault is uniquely favoured, even amongst competing stacking faults that also preserve the stoichiometric local environment.

Electronic structure analysis suggests that these (001) stacking faults are electrically benign, i.e. do not introduce deep-level trap states within the fundamental band gap. Furthermore, our results indicate that band alignments remain largely unperturbed, suggesting that these intrinsic planar defects are not detrimental to minority carrier diffusion in $Zn_3P_2$ devices.

While these static models indicate a negligible direct electronic impact, we acknowledge that indirect mechanisms, such as the effects of strain or the role of stacking faults as nucleation sites for point defects, requires further investigation. Nevertheless, by establishing the benign electronic nature of these ubiquitous defects, this work redirects the search for efficiency-limiting mechanisms toward defect complexes and interface engineering.

# Acknowledgements

The authors acknowledge funding from the European Innovation Council and SMEs Executive Agency (EISMEA) under grant agreement No 101046297 (Pathfinder Project SOLARUP). The authors thank the colleagues of the SOLARUP consortium for fruitful scientific discussions and for their critical review of the manuscript. ICN2 is supported by the Severo Ochoa program from Spanish MCIN / AEI (Grant No.: CEX2021-001214-S) and is funded by the CERCA Programme / Generalitat de Catalunya. ICN2 acknowledges funding from Generalitat de Catalunya 2021SGR00457. Part of the present work has been performed in the framework of Universitat Autònoma de Barcelona Materials Science PhD program. Authors acknowledge the use of instrumentation as well as the technical advice provided by the Joint Electron Microscopy Center at ALBA (JEMCA). ICN2 acknowledges funding from Grant IU16-014206 (METCAM-FIB) funded by the European Union through the European Regional Development Fund (ERDF), with the support of the Ministry of Research and Universities, Generalitat de Catalunya. ICN2 is founding member of e-DREAM.[85]

The authors gratefully acknowledge the computing time provided to them on the high-performance computers Noctua 2 at the NHR Center PC2. These are funded by the Federal Ministry of Education and Research and the state governments participating on the basis of the resolutions of the GWK for the national high-performance computing at universities (www.nhr-verein.de/unsere-partner). Crystal structures were visualized with VESTA. We made use of the sumo package for post-processing and analysis of the DFT calculations.

The authors from Lund University acknowledge support from NanoLund and the Lund Nano Lab (myfab Lund) in the growth of the analysed samples.

# Supplementary Information

# Unravelling the Role of Stacking Disorder on the Optoelectronic Properties of $Zn_3P_2$

Francesco Salutari[1,†], Nico Kawashima[2,3,†], Aidas Urbonavicius[4], Helena Rabelo Freitas[1], Raphael Lemerle[5], Thomas Hagger[5], Kimberly A. Dick[4], Anna Fontcuberta i Morral[5,6], Simon Escobar Steinvall[4], Maria Chiara Spadaro[1,7,8], Silvana Botti[2,*], Jordi Arbiol[1,9,*]

[1]Catalan Institute of Nanoscience and Nanotechnology – ICN2 (CSIC and BIST), Campus UAB, Bellaterra, Barcelona, 08193, Catalonia, Spain
[2]Research Center Future Energy Materials and Systems and Interdisciplinary Centre for Advanced Materials Simulation, Faculty of Physics and Astronomy, Ruhr University Bochum, Universitätsstraße 150, 44801 Bochum, Germany
[3]Institute of Condensed Matter Theory and Optics, Friedrich-Schiller-Universität Jena, Max-Wien-Platz 1, 07743 Jena, Germany
[4]Center for Analysis and Synthesis and NanoLund, Lund University, Box 124, 221 00 Lund, Sweden
[5]Laboratory of Semiconductor Materials, Institute of Materials, School of Engineering, Ecole Polytechnique Fédérale de Lausanne, 1015, Lausanne, Switzerland
[6]Institute of Physics, EPFL, Route Cantonale 1, Lausanne, 1015, Vaud, Switzerland
[7]Physics and Astronomy Department "Ettore Majorana" (DFA), Catania University, Via S. Sofia 64, Catania, 95123, Sicily, Italy
[8]IMM-CNR, Catania University Building, Via S. Sofia 64, Catania, 95123, Sicily, Italy
[9]ICREA, Pg. Lluis Companys 23, Barcelona, 08010, Catalonia, Spain

†Equal Contribution
*Corresponding authors

Figure SI1 and SI2, we report HAADF-STEM images and corresponding dark-field images obtained via FFT-based filtering of different $Zn_3P_2$ samples. Planar defects manifest as parallel dark-contrast lines which lye within the (001) plane.

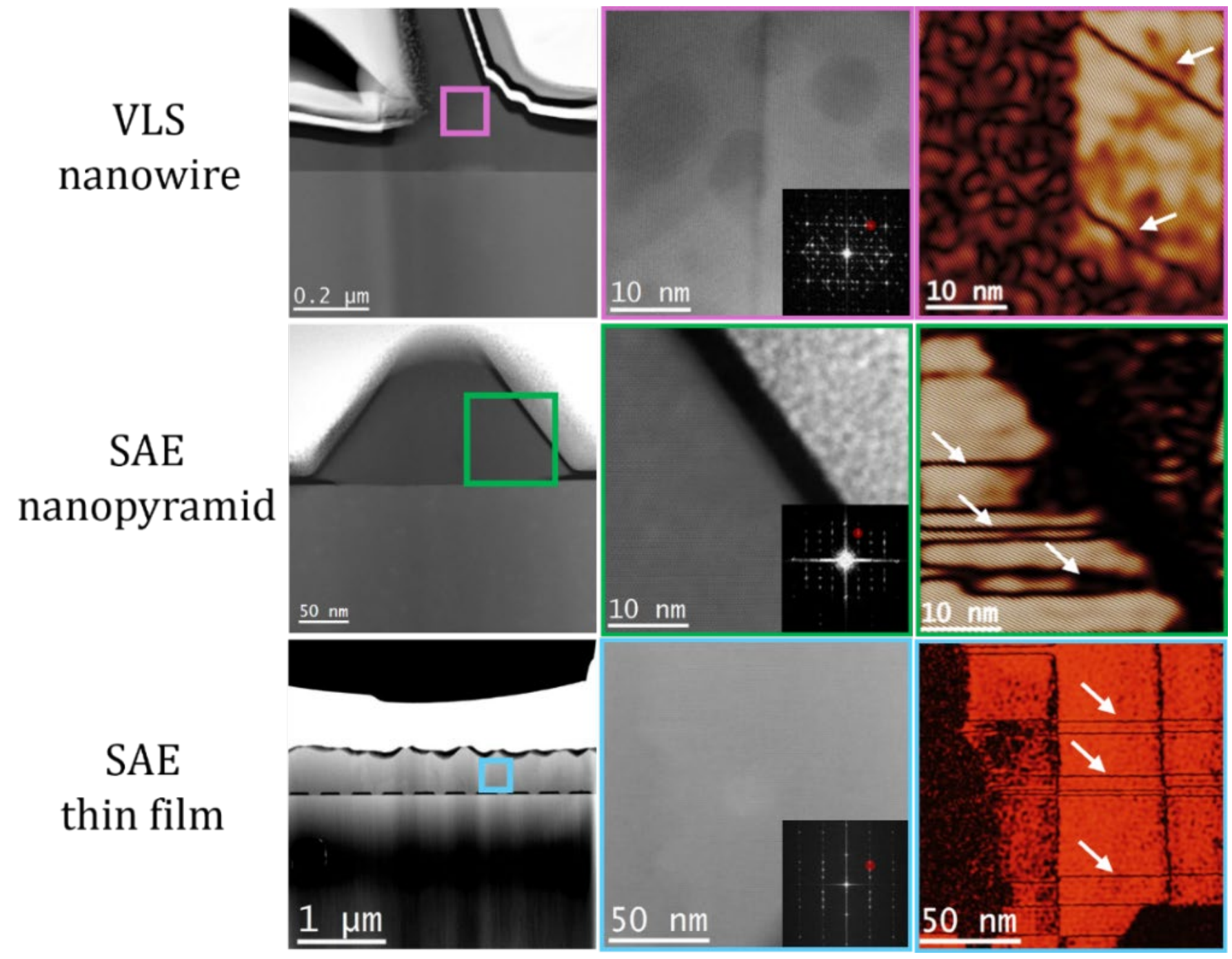

*Figure SI 1: Planar defects are highlighted as parallel dark-contrast lines (indicated by the white arrows) by means of FFT-based filtering in different types of $Zn_3P_2$ structures grown on InP. The regions considered here are all oriented along the <100> zone axis and planar defects are parallel to the (001) plane.*

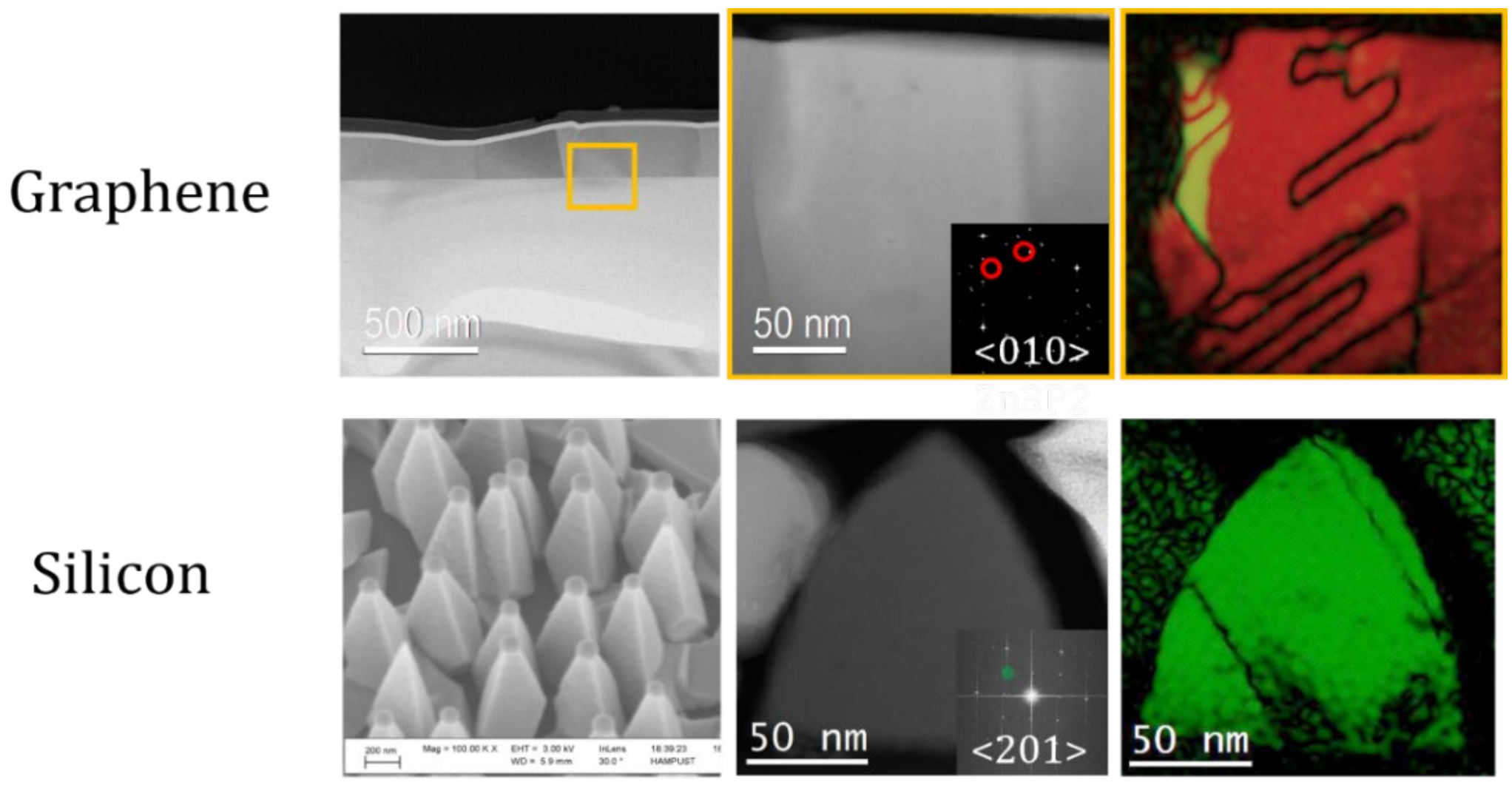

*Figure SI 2: Dark-contrast lines appear in $Zn_3P_2$ grown on top of Silicon and Graphene. This suggests that the interfacial epitaxial relation with the substrate does not influence the appearance of planar defects. The sample grown on Si is imaged along a different zone axis, namely the <201>. In Figure SI 4, we show that such contrast is consistent with the proposed 3D atomic model of stacking faults.*

In Figure SI3, we report atomic resolution HAADF-STEM and iDPC-STEM images of a defective region of $Zn_3P_2$ oriented along the [100] zone axis. The 3D atomic model based on the displacement vector $\boldsymbol{R}$ = [0, ½, 0] is overlaid to the iDPC-STEM image. The position of the light P atoms is consistent with the model.

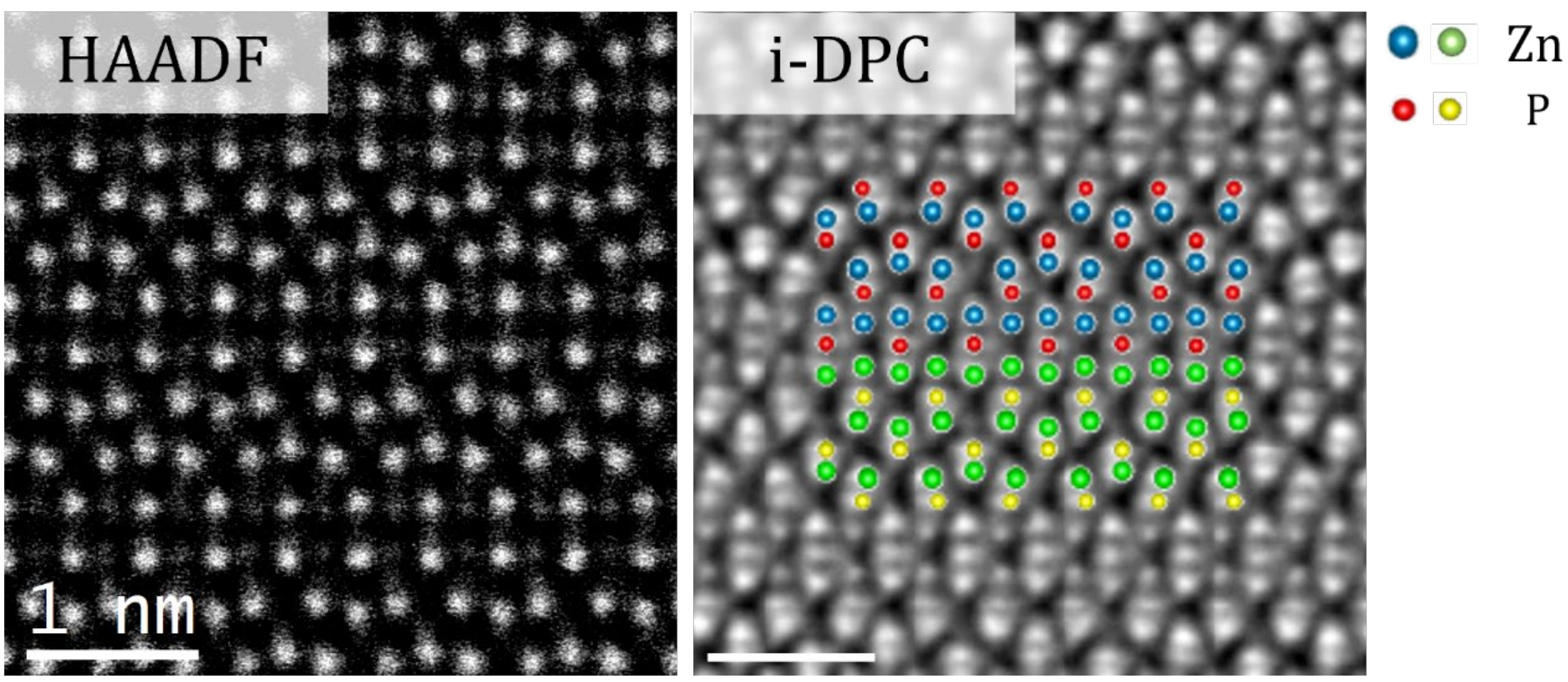

*Figure SI 3: 3D atomic model of the faulted structure superimposed to the i-DPC STEM image of $Zn_3P_2$ oriented along the <100> zone axis. The position of the light P atoms is consistent with the model.*

In Figure SI4 we show additional zone axis that are consistent with the 3D atomic model described in the main text. Such orientations are not parallel to the (001) plane, thus the defect plane does not manifest as a sharp interface

## SI5 Effect of Crystal Symmetry on Displacement Vector Identification

The extinction criteria used to interpret the diffuse streaking observed in the experimental SAED pattern do not fully resolve the ambiguity in the displacement vector, as the data are equally consistent with both ***R*** = [0, ½, 0] and ***R*** = [½, ½, 0]. This ambiguity arises from the tetragonal symmetry of $Zn_3P_2$, for which the [100] and [010] directions are equivalent and therefore indistinguishable in their two-dimensional projections. Furthermore, the presence of a local defect may locally break the fourfold rotational symmetry of the crystal.

This point can be clarified by considering the atomic models shown below. The model in Figure SI5a is obtained using the displacement vector ***R*** = [0, ½, 0]. As discussed in the main text, this model correctly reproduces the experimental image contrast along the [100] zone axis (as well as along other zone axes; see Figure 2 and Figure SI4). In addition, the contrast associated with the planar defect remains consistent with the fourfold symmetry of the crystal through the **g·R** invisibility criterion, according to which the defect becomes invisible when **g·R** =0. Along the [010] direction, the diffraction vectors are of the form **g**=(h0l), yielding **g·R** =0, and therefore the defect contrast disappears.

In contrast, the fourfold rotational symmetry is not preserved for the displacement vector ***R*** = [½, ½, 0], as illustrated by the model in Figure SI5b. In this case, the symmetry operation disrupts the long-range chemical ordering and generates energetically unfavorable nearest-neighbor atomic configurations across the interface.

The position of the faulted plane can be shifted by one atomic layer, as shown in Figure SI5c. Under these conditions, both displacement vectors, ***R*** = [0, ½, 0] and ***R*** = [½, ½, 0] preserve the fourfold rotational symmetry.

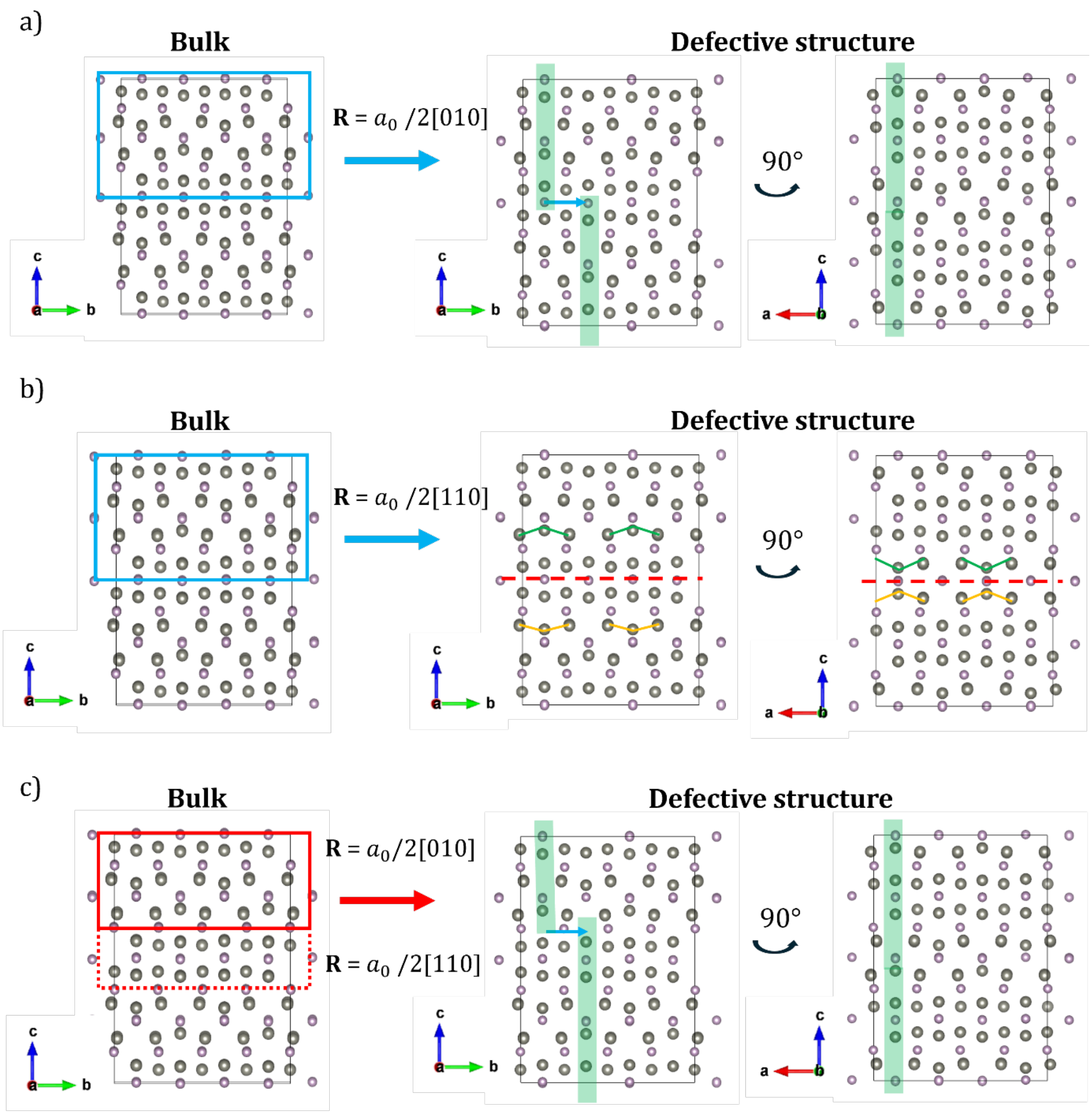

*Figure SI 5: 3D atomic models of the defective $Zn_3P_2$ crystal structure obtained by rigid translations applied to the pristine bulk structure. a) The atoms within the blue box are displaced along the [010] direction by half the unit cell parameter. Such operation preserves the 4-fold rotational symmetry hiding the defect when imaging the crystal along the [010] direction. b) The fault plane is the same as in a) but now the displacement vector is of the form* ***R*** *= a/2[110]. The operation does not preserve the rotational symmetry as unfavoured nearest-neighbour configurations appear across the interface. c) The 4-fold rotational symmetry is preserved for both* ***R*** *= a/2[110] and* ***R*** *= a/2[010] if the fault plane is shifted by one atomic layer.*

In Figure SI6, we report the HAADF-STEM image, along with the corresponding 3D atomic model, of a region of the crystal with ordered stacking sequences. The ordering is not preserved for more than two periods.

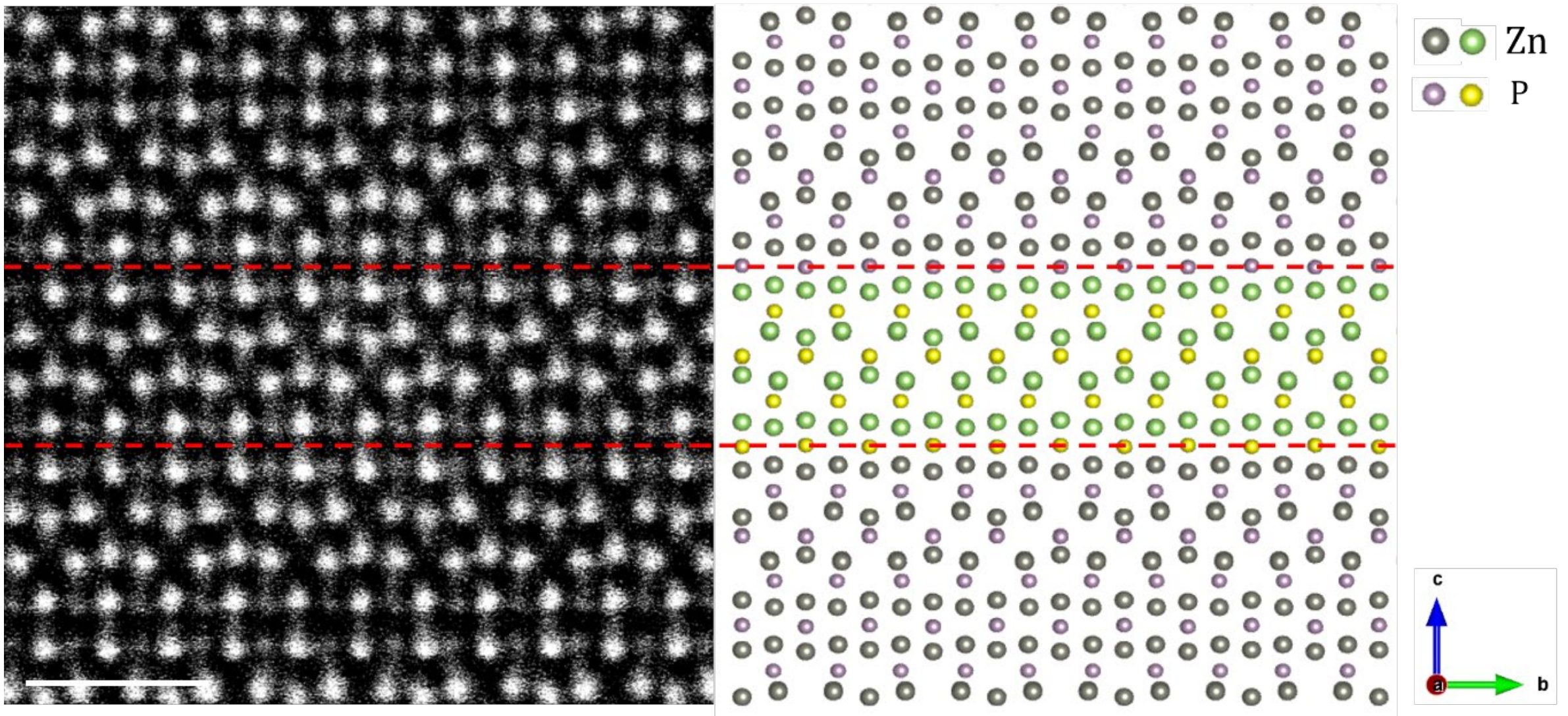

*Figure SI6: HAADF image and corresponding 3D atomic model of an ordered stacking sequence. Experimentally we have not observed the ordered sequence repeating for more than two periods. Scale bar is 1nm.*

In Figure SI7, we simulate the SAED patterns of ordered defective structures with long-range ordering and compare to the diffraction pattern of the pristine bulk structure. As the period increases, the number of superlattice-like reflections increases. The limit scenario is represented by the experimental observation, where the absence of short- and long- range ordering produces diffuse streaking in the diffraction pattern.

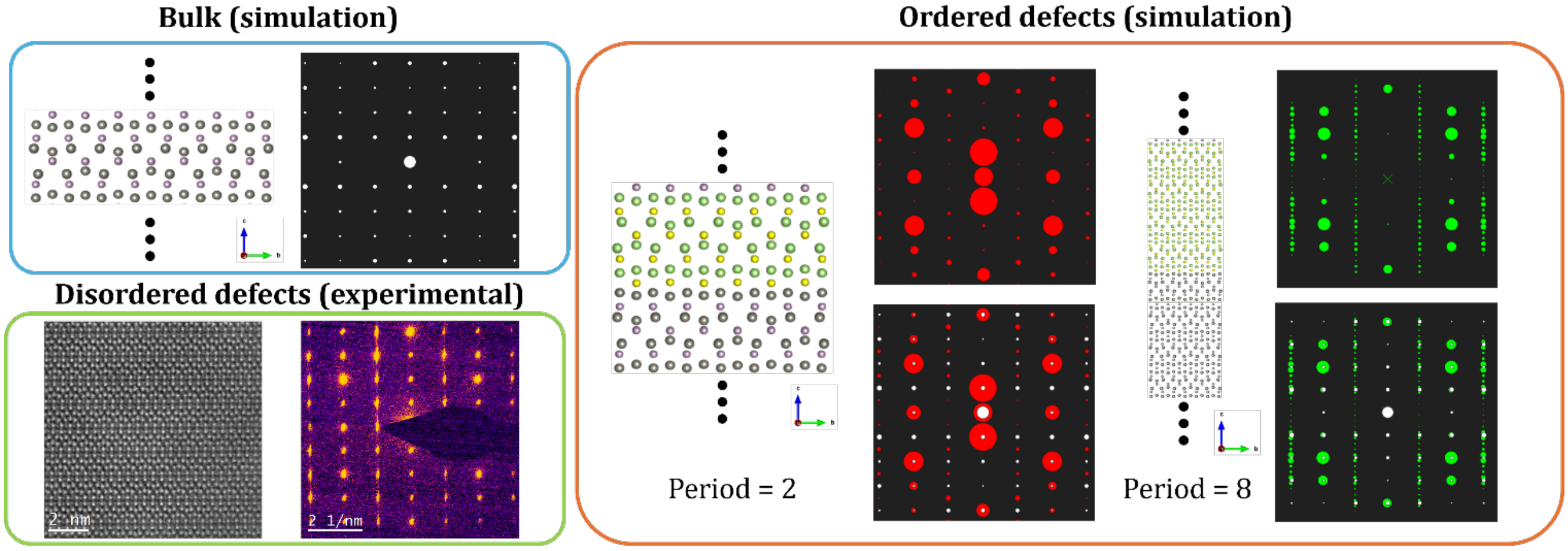

*Figure SI7: Simulated SAED diffraction patterns of bulk $Zn_3P_2$, experimental disordered structure and defective structures with long-range ordering. The ordered structures display extra superlattice-like reflections symmetric to the bulk Bragg spots of type (hkl) with h = odd. To highlight the extra reflections and their relation with the bulk SAED pattern, we superimposed the SAED pattern of the ordered structures with the SAED pattern of the bulk structure. The number of extra reflections scales linearly with the period of the faulted stacking sequence. In contrast, the experimentally observed $Zn_3P_2$ structure lacks both short-range and long-range ordering; consequently, the effective stacking period tends toward infinity, resulting in diffuse streaking features in the experimental SAED pattern.*

## SI8 Combinatorial exploration and ab initio investigation of alternative stacking faults

To evaluate the formation of stacking faults in $Zn_3P_2$ without empirical bias, we performed an experiment-agnostic combinatorial exploration of the configuration space. Given the complexity of the $Zn_3P_2$ unit cell, we had to rely on physical constraints to reduce the vast number of theoretically possible planar defects to a small subset of physically plausible configurations. These constraints are motivated by the known defect chemistry of the material, specifically the prevalence of zinc-related point defects over phosphorus-related ones.

First, since zinc interstitials ($Zn_i$) and vacancies ($V_{Zn}$) exhibit significantly lower formation energies than their phosphorus counterparts, planar defects were modelled as stacking faults (SF) of the Zn sublattice, i.e., as variations in the ordering of zinc “vacancies” compared to a fluorite structure. Second, we enforced a strict coordination constraint based on the pristine $P4_2/nmc$ phase, where each P anion is coordinated by exactly six Zn cations. This six-fold coordination aligns with a model where Zn acts as a +2 cation and P as a −3 anion.

By preserving the P-sublattice and enforcing the stated coordination condition during the redistribution of Zn atoms, a combinatorial search was performed within the tetragonal unit cell. This exploration revealed that the configuration space collapses into only nine unique structures. Each may be described as a specific periodic stacking of eight different Zn (001) layers, labeled A through H as illustrated in Figure SI8. These nine structures can be split into two groups: The first group comprises the pristine bulk stacking ABCD, as well as three SF composed solely of the layers present in the pristine unit cell,

ABDC, ABBA, and ABAB,

where the first two lead to a subsequent formal reversal of the stacking order. The second group consists of five SF formed by replacing adjacent Zn sublayer pairs AB or CD (but not BC or DA) in the pristine ABCD sequence with two layers selected from within the set E through H,

ABEF, ABEG. ABGH, ABEE and ABGG.

Notably, single-layer substitutions are not possible as they violate the required P coordination condition.

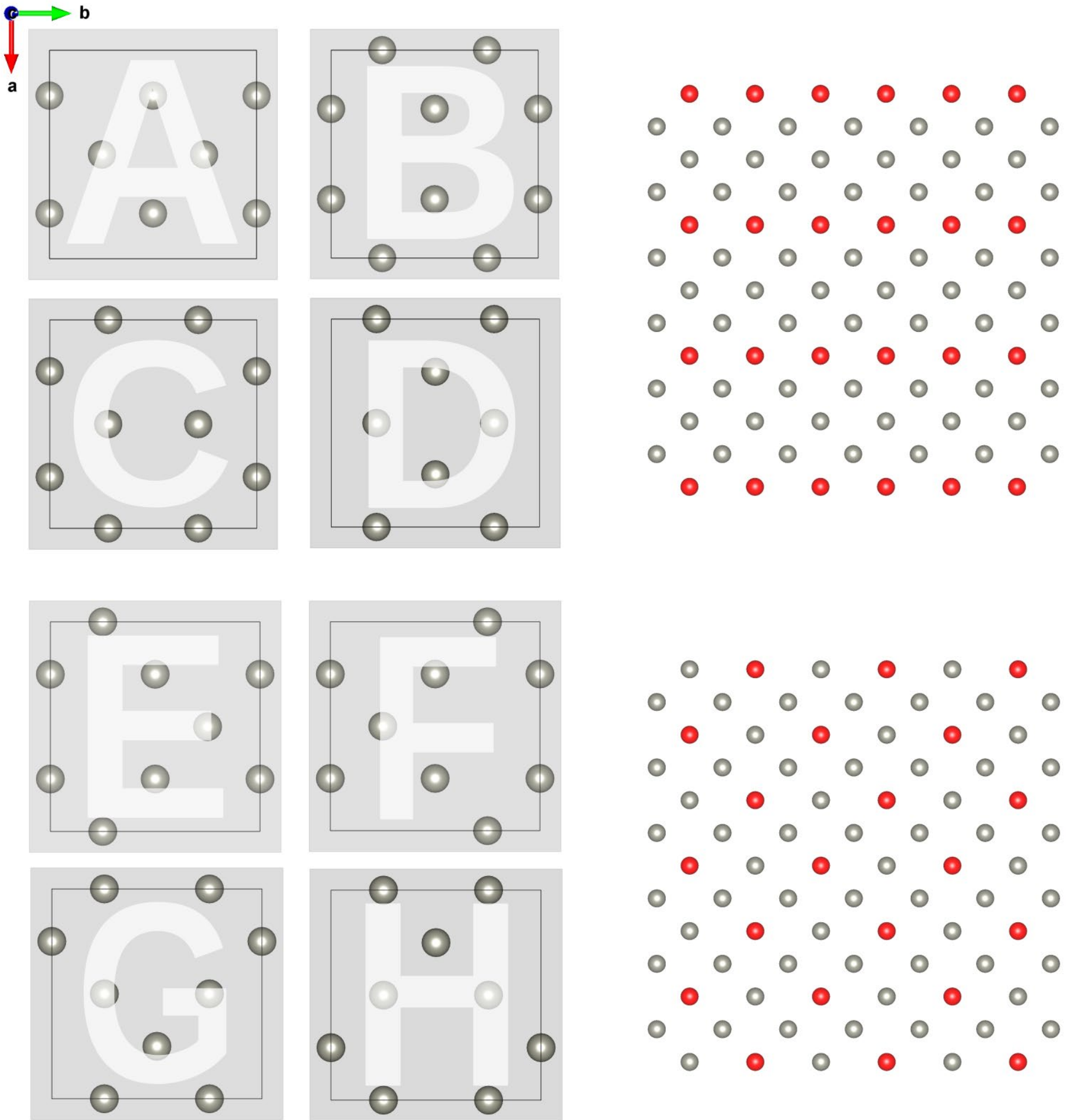

***FIGURE SI8:*** *Illustrations and definitions of unique Zn (001) layer configurations. (Left) The eight unique layers are partitioned into two sets: the pristine types A to D and the alternative types E to H. (Right) Reference vacancy distributions for each set, where Zn atoms and "vacancies" are denoted by grey and red spheres, respectively. Within each set, individual layers are symmetry-equivalent and can be mapped onto one another via in-plane translations or 90° rotations.*

The formation energies for various configurations were computed using machine-learned interatomic potentials (MLIP) and validated via density functional theory (DFT). Given the high computational cost of *ab initio* methods, DFT calculations were restricted to the three SF consisting of layers A to D. Comparison to the formation energies obtained from MLIP demonstrate that relying on MLIP accuracy is sufficient for evaluating the other SF. As summarized in Table SI8, the ABDC stacking fault is uniquely favorable, with a formation energy of only 2–3 mJ/m$^2$, which is significantly lower than the configuration with the next-lowest formation energy, ABGG, at 171 mJ/m$^2$. This significant energetic preference explains why the ABDC fault is the primary planar defect observed experimentally, as discussed in the main text.

| Stacking fault (representative sequence) | Formation energy $E_f$ in mJ/m$^2$ (MLIP) | Formation energy $E_f$ in mJ/m$^2$ (DFT) |
|---|---|---|
| ABCD **ABDC** BADC | 3.(8) | 2.(5) |
| ABCD **ABGG** ABCD | 17(1) | |
| ABCD **ABGH** ABCD | 22(1) | |
| ABCD **ABBA** DCBA | 24(3) | 22(4) |
| ABCD **ABEG** ABCD | 24(3) | |
| ABCD **ABAB** CDAB | 33(8) | 33(8) |
| ABCD **ABEE** ABCD | 45(0) | |
| ABCD **ABEF** ABCD | 53(5) | |

**TABLE SI8:** Computed stacking fault formation energies $E_f$ for all eight structures investigated. Results are provided for both, via machine-learned interatomic potentials (MLIP) and via density functional theory (DFT), where applicable. Stacking sequences are denoted such that bold characters indicate the specific stacking fault region, while non-bold characters represent the surrounding bulk-like sequences that are periodically repeated.

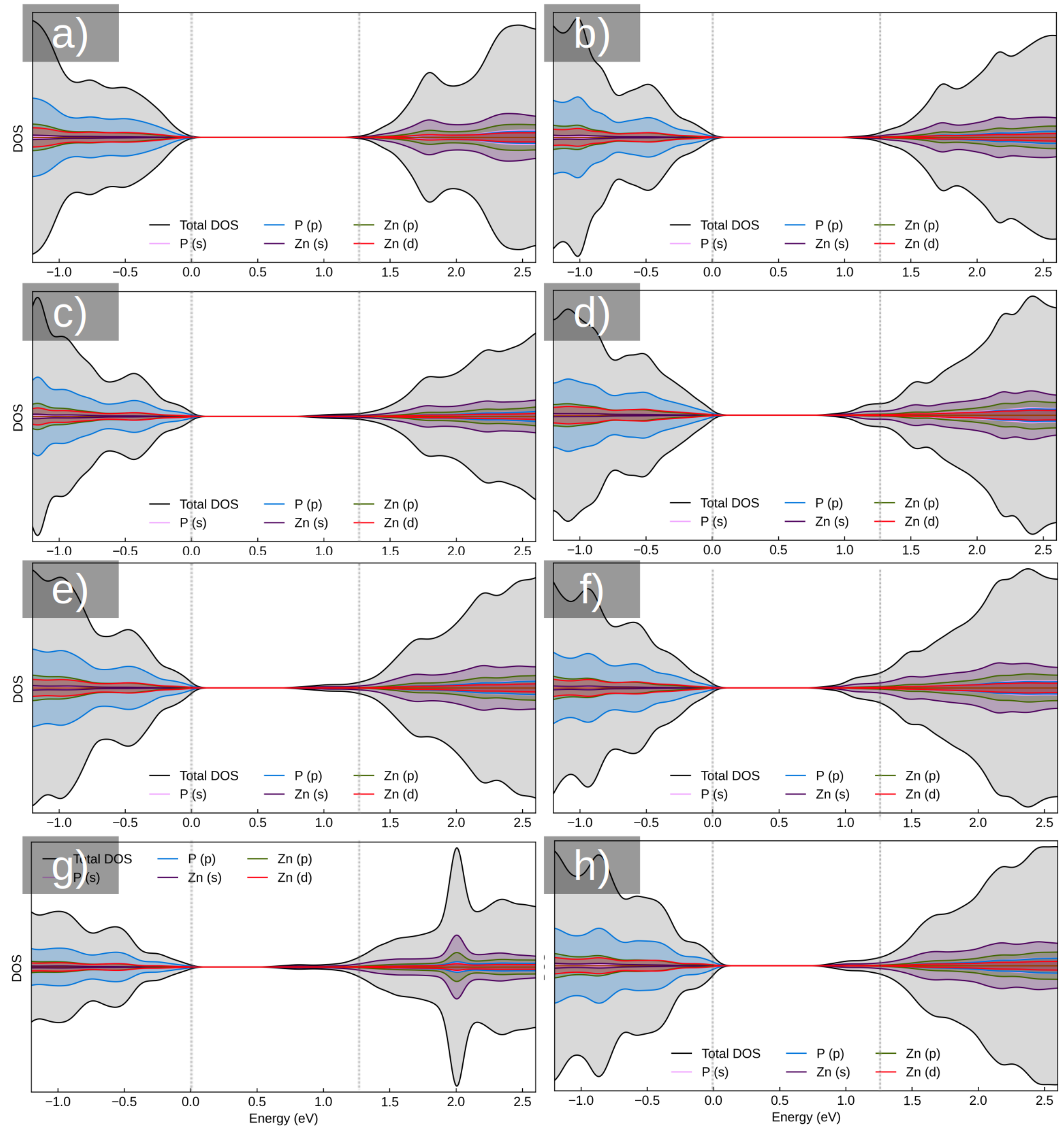

**FIGURE SI9:** Density of states (DOS) of the ideal $Zn_3P_2$ crystal (a) compared with the DOS of structures containing various stacking faults (b–h). The stacking sequences corresponding to panels (b–h) are ABGG, ABGH, ABBA, ABEG, ABAB, ABEE, ABEF, respectively. To remove sampling artifacts, a post-processing Gaussian broadening was applied with a standard deviation of 0.05 eV. Dashed grey lines denote the valence band maximum and conduction band minimum of the ideal crystal for better comparison.